\documentclass{ws-ijmpd}
\usepackage[super,compress]{cite}
\usepackage{cprotect,hyperref,amsmath,amsfonts,amssymb,bm}
\usepackage{calrsfs}
\usepackage{physics}
\usepackage{graphicx}
\makeatletter
\newcommand*{\rom}[1]{\expandafter\@slowromancap\romannumeral #1@}
\makeatother
\let\oldhat\hat

\renewcommand{\hat}[1]{\oldhat{\mathbf{#1}}}
\usepackage[super]{nth}
\begin{document}
	\markboth{Amir A. Khodahami and Azizollah Azizi}
	{Holography and the Page curve of an evaporating black hole}
	%%%%%%%%%%%%%%%%%%%%% Publisher's Area please ignore %%%%%%%%%%%%%%%
	%
%	\catchline{}{}{}{}{}
	%
	%%%%%%%%%%%%%%%%%%%%%%%%%%%%%%%%%%%%%%%%%%%%%%%%%%%%%%%%%%%%%%%%%%%%
	\title{Holography and the Page curve of an evaporating black hole}
	\author{Amir A. Khodahami}	
	\address{Department of Physics, Shiraz University, Shiraz 71949-84795, Iran\\a.khodahami@shirazu.ac.ir}
	\author{Azizollah Azizi}
	\address{Department of Physics, Shiraz University, Shiraz 71949-84795, Iran\\azizi@shirazu.ac.ir}
	\maketitle
%	\begin{history}
%		\received{Day Month Year}
%		\revised{Day Month Year}
%	\end{history}
%%%%%%%%%%%%%%%%%%%%%%%%%%%%%%%%%%%%%%%%%%%%%%%%%%%%%%%%%%%%%
		\begin{abstract}
		\textbf{Abstract:} The Page curve exhibits a sharp peak at the Page time corresponding to a phase transition from an empty quantum extremal surface to a non-empty one. This study delves into the impact of this phase transition on the informational content of black hole radiation through constructing a smooth Page curve which represents a gradual transition instead. We utilize replica trick, apply holography in a specific form, and introduce mathematical identities inherent to the curve's smoothness. Our results suggest that a gradual transition could lead to a significant increase in the informational content of the radiation.
	%%%%%%%%%%%%%%%%%%%%%%%%%%%%%%%%%%%%%%%%%
		\end{abstract}
%	\ccode{PACS numbers:}
%%%%%%%%%%%%%%%%%%%%%%%%%%%%%%%%%%%%%%%%%%%%%%%%%%%%%%%%%%%
\section{Introduction}\label{Sec__Intro}
\begin{figure}[t]
	\centering
		\includegraphics[width=\linewidth]{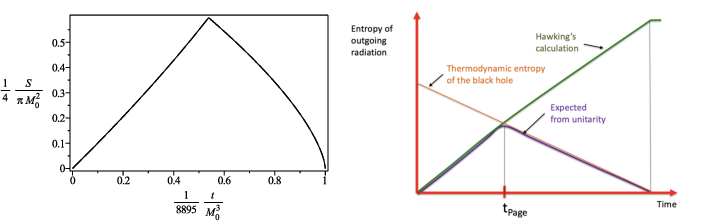}
	\caption{\label{f__Page} Left: The curve that Page suggested for entanglement entropy of the radiation emitted from a black hole initially in a pure state, from \cite{Page2013}. This curve is produced by considering the thermodynamic entropies of the radiation and the black hole as functions of time, and choosing the one that is less. Right: A schematic visualization of Page's procedure in obtaining the Page curve, from \cite{Almheiri2021}. As is shown, the true curve for entanglement entropy of the radiation is expected to follow the thermodynamic entropy of the radiation at first. However, when reaching the Page time, it should switch to follow the thermodynamic entropy of the black hole. Obviously this curve is consistent with unitarity.}
\end{figure}

In 1975, Hawking studied quantum field theory in curved spacetime and showed that black holes radiate and hence lose their masses \cite{Hawking1975,Page1976}. This is in accord with the former obtained temperature of black holes, $T=\kappa/2\pi$, where $\kappa$ is black hole's surface gravity \cite{Bekenstein1972,Bardeen1973,Page2005}. Then he proposed a paradox which was concerning the conservation of quantum information and predictability \cite{Hawking1976}. In fact, he showed that the black hole formation and evaporation process cause pure states to evolve into mixed states. Hence it destroys quantum information and is not a unitary process. A quantity called ``von Neumann entropy'' or ``fine-grained entropy'' or ``entanglement entropy'' is used to express the information loss amount quantitatively, which is defined as $S_R=-\Tr\rho_R\ln\rho_R$, where $\rho_R$ is the density matrix of the radiation. One would expect this quantity to rise at initial stages of the evaporation due to the entanglement between the radiation emitted and the remained black hole. However, in the case of starting from a black hole in a pure state, it should fall down after a time and eventually hit zero for the process to be unitary. Such a curve is suggested by Page by considering the thermodynamic entropies of radiation and black hole and switching between them as is shown in Fig.~\ref{f__Page} \cite{Page1983,Page1993,Page2013}. According to the Page curve, although the entanglement entropy of radiation follows Hawking's calculation at first, after a time called Page time---when the thermodynamic entropies of radiation and black hole become equal---it turns to follow the thermodynamic entropy of the black hole, which is given by the Bekenstein-Hawking formula, $S=\text{Area}/4G$ (there are several ways to obtain this formula microscopically \cite{Strominger1996,tHooft1985,Srednicki1993,Bombelli1986,Bena2022,Balasubramanian2022}).
\par Much attempts made to resolve the paradox and obtain Page curve which finally led to the so-called ``island formula'' for entanglement entropy of the radiation \cite{Almheiri2019,Almheiri2020,Almheiri2020a,Penington2020}. One can see, in the language of path integral, that Hawking's calculations neglect some paths which although are tiny at initial stages of the evaporation process, become dominated at final ones. The following relation is obtained for entanglement entropy in gravitational systems
\begin{equation}\label{Eq__EE_QES}
	S=\text{min}_X\left\{\text{ext}_X\left[\frac{\text{Area}(X)}{4G}+S_{\text{semi-cl}}\left(\Sigma_X\right)\right]\right\},
\end{equation}
where $X$ is a codimension-2 surface and $\Sigma_X$ is the region bounded by $X$ and the surface from where we can almost assume that spacetime is flat, called cutoff surface\footnote{Determining the precise location of the cutoff surface is challenging according to the long-range nature of gravity \cite{Almheiri2021}.}. Moreover, $S_{\text{semi-cl}}$ is the von Neumann entropy of quantum fields on $\Sigma_X$, appearing in semiclassical description \cite{Ryu2006,Ryu2006a,Hubeny2007,Engelhardt2015,Faulkner2013} (a good review is presented in \cite{Almheiri2021}). Any surface that extremizes the expression in the square brackets is called ``Quantum Extremal Surface'' (QES). It can be shown that there are two such surfaces present in the evaporation process, one with vanishing area and the other tending to the event horizon of the black hole from below. At the initial stages of the evaporation process, the first one can be found to be the minimizing QES which represents the no-island solution. However, after a time (Page time) the other one becomes dominated, representing the so-called island solution---due to the presence of a non-vanishing surface inside the black hole.
\par
The precise state of the radiation is determined by the details of the evaporation process that made it. So, although gravitational interactions can be ignored in the radiation region $\Sigma_\text{rad}$ (from the cutoff surface all the way to infinity), the precise state of the radiation can only be determined via gravity. Thus one should again make use of Eq.~\eqref{Eq__EE_QES} for truly computing the entanglement entropy of the emitted radiation. It should be noticed that the term $\Sigma_X$ for the radiation consists of: i) $\Sigma_\text{rad}$ and ii) $\Sigma_\text{island}$---the region inside the island. According to the literature, ``island'' is a region that is disconnected from the radiation region but contributing in its entanglement entropy in the form of Eq.~\eqref{Eq__EE_QES}. Having the initial black hole in a pure state causes the QESs of the radiation to be identical with those of the black hole. In this work, we assume the initial black hole to be in a pure state, so that we could take the entanglement entropies of the radiation and the black hole to be identical.
\par The thermodynamic entropy of a black hole is given by the Bekenstein-Hawking formula, $S=A/4G$. It scales with the radius squared and not cubed. This property is well understood in AdS/CFT correspondence---the most successful realization of the holographic principle---as there is a bijective isometry between the two theories in two spaces with different dimensions: a theory of gravity in a $(d+1)$-dimensional AdS bulk and a quantum field theory with conformal symmetry living on its $d$-dimensional boundary \cite{Hooft1993,Maldacena1997,Witten1998}. Moreover, AdS/CFT correspondence plays a central role in derivation of Eq.~\eqref{Eq__EE_QES}. In fact, this relation is based on the so-called RT/HRT/EW bulk reconstruction, which itself is studied and formulated in the light of AdS/CFT correspondence \cite{Ryu2006,Ryu2006a,Hubeny2007,Engelhardt2015,Rangamani2016}\footnote{The information paradox and its resolution is also worked out in de Sitter and asymptotically flat spacetimes \cite{Marolf2021,Marolf2021a,Geng2021,Arefeva2021,Hartman2020,Hashimoto2020,Wang2021,Gautason2020}.}. Hence the holographic formula, $S=A/4G$, which controls the limiting behavior of $S$ for $r\rightarrow 0$, or equivalently for $t\rightarrow t_{\text{evap}}$, is quite intrinsic in AdS/CFT correspondence. So the strict switching between the two thermodynamic entropies that Page did, is now going to be done by $\text{min}_X$ in Eq.~\eqref{Eq__EE_QES}.
\par The Page curve exhibits a singular (breaking) point at the Page time, arising from the interplay between the thermodynamic entropies of the black hole and the radiation. Specifically, within the context of the replica trick in the Euclidean framework, this phenomenon corresponds to a phase transition from a disconnected Euclidean topology (the vanishing QES) to a connected one (the non-vanishing QES) \cite{Penington2022a}. It is worth noting that a realistic Page curve will not exhibit a perfectly sharp peak at the Page time. Even in typical phase transitions, such as the liquid-gas phase transition, Finite Size Effects come into play due to the finite number of particles ($N$). Accordingly, for the Page curve, there exist corrections of $\mathcal{O}(1/\sqrt{G})$ near the transition point \cite{Marolf2020,Dong2020}. Moreover, it is shown in \cite{Akers2021} that the QES prescription leads to contradictions, which can be resolved by introducing corrections of $\mathcal{O}(1/G)$ (the same order as the dominant contribution in the QES prescription). They have refined the QES prescription by recognizing that different parts of the system’s state may have undergone the phase transition or not. It is different from the QES prescription in which the entire state has to be on one side of the phase transition or the other. Hence the strict transition in the QES prescription is smoothed by some $\mathcal{O}(1/G)$ correction.
\par Our investigation focuses on the impact of this phase transition on the informational content of black hole radiation. To be more concrete, we aim to construct a smoothed-out Page curve corresponding to a gradual transition and explore its implications for the informational content of the emitted radiation. Our results indicate that a gradual transition may result in a substantial increase of around 50\% in the informational content of the radiation at the Page time---$S^{\text{grad}}_{\text{max}}/S^{\text{sharp}}_{\text{max}}\sim 0.5$. It is worth noting that, in this work, we use replica trick and consider holography in form of the following assumptions to progress our calculations, while controlling the behavior of $S$ in the limit of $r\rightarrow 0$, or equivalently $t\rightarrow t_{\text{evap}}$: (1) In the replicated system, the entanglement entropy of each black hole scales with its area; (2) The black hole's entanglement entropy will not exceed the Bekenstein-Hawking entropy \cite{Flanagan1999,Bousso2003,Bekenstein2005,Casini2008,Chen2008,Horwitz2022}. Additionally, we introduce and use certain mathematical identities under the assumption that the intended entanglement entropy is smooth and hence expandable in terms of some boundary deformations. Obviously, this work is not going to give a resolution to the information paradox. Rather it proposes a method to build a smooth version of the Page curve by making use of the holographic assumptions, alongside some conditions necessary for the curve to be smooth.
\par This paper is organized as follows: In Sec.~\ref{Sec__Replica}, we provide a concise overview of the replica trick and discuss its significance in deriving the true curve for entanglement entropy---the Page curve. In Sec.~\ref{Sec__Shape_Dependence}, we establish some mathematical identities which will help us a lot in obtaining the smooth version of the Page curve. In Sec.~\ref{Replica__EE}, we show how the desired curve can be obtained by making use of the holographic assumptions and the mathematical identities. Moreover, some comments on the higher-order modifications to the curve are discussed in subsection \ref{Sec__higher_replicas}. We compare our derived curve with a smooth fit to the original Page curve, as detailed in subsection \ref{Sec__Num_Ana}. Finally conclusions are presented in Sec.~\ref{Sec__Summ}. 
\section{Replica trick}\label{Sec__Replica}
Replica trick is a mathematical tool that helps us to calculate entanglement entropy of a system $\mathcal{A}$, with its environment $\bar{\mathcal{A}}$. Let assume that the system and its environment as a whole (call it $\mathcal{A}\bar{\mathcal{A}}$) has a wave function $\ket{\Psi}$ which itself has an expansion in terms of some eigenstates $\ket{\psi}_i\ket{\bar{\psi}}_i$, of $\mathcal{H}\times\bar{\mathcal{H}}$, with $\mathcal{H}$ and $\bar{\mathcal{H}}$ being Hilbert spaces of $\mathcal{A}$ and $\bar{\mathcal{A}}$ respectively
\begin{equation}
	\ket{\Psi}=\sum_{i} C_i\ket{\psi}_i\ket{\bar{\psi}}_i.
\end{equation}
One can obtain the density matrix $\varrho$, of $\mathcal{A}\bar{\mathcal{A}}$ as
\begin{equation}
	\varrho=\ket{\Psi}\bra{\Psi}.
\end{equation}
Taking a trace over the environment, $\bar{\mathcal{A}}$, will lead to the density matrix, $\rho$, of the system, $\mathcal{A}$
\begin{equation}
	\rho=\Tr_{\bar{\mathcal{A}}}\left(\varrho\right).
\end{equation}
We have $\Tr_{\mathcal{A}}(\rho)=1$ due to the normalization of $\ket{\Psi}$. The entanglement entropy, which is a measure of how much the system and its environment are entangled to each other, is given by
\begin{equation}\label{Eq__EntEnt}
	S=-\Tr_{\mathcal{A}}\left(\rho\ln\rho\right).
\end{equation}
As is shown by the index $\mathcal{A}$, the trace is taken over the system. From now on, we do not explicitly write the index $\mathcal{A}$ for briefness; but bear it in mind.
\par One can equivalently obtain the entanglement entropy by
\begin{equation}
	S=-\frac{\partial}{\partial n}\ln(\Tr\rho^n)\biggl|_{n\rightarrow1},
\end{equation}
which for a normalized wave function (i.e., $\Tr\rho=1$) can also be written as
\begin{equation}
		S=-\frac{\partial}{\partial n}\Tr\rho^n\biggl|_{n\rightarrow1}.
\end{equation}
Hence to obtain the entanglement entropy, we first consider $n$ replicas of the system and find $\rho^n$. Then we take a trace (and a natural logarithm) followed by a derivative with respect to $n$. Finally, we perform the limit $n\rightarrow1$ and put a minus sign.
\par It should be noticed that the number of replicas, $n$, should approach 1 continuously in the limit. Thus, an analytic continuation of $n$ is needed for the limit to be well-defined. This analytic continuation sometimes runs calculations into trouble (a proposal is presented in \cite{Wu2023} to tackle such difficulties using ``deep learning''). However, there is a strict reason to make use of the replica trick in some cases, which is explained in the following subsection.
\subsection{Why replica trick?}\label{Sec__whyrt}
As explained before, replica trick is a mathematical tool that would be used to derive the true curve for entanglement entropy of an evaporating black hole---the Page curve. In fact, it helps to take all the terms participating in the entanglement entropy into account. When a black hole evaporates, it undergoes a transition from an initial state to a final state. All the paths emanating from the initial state and ending in the final one should be considered for obtaining the true result. However, some paths might be difficult to see when we perform direct calculations by the means of Eq.~\eqref{Eq__EntEnt}. As a matter of fact, such paths will appear in calculations through some non-trivial topologies when using replica trick. As an example, for the case of $n=2$ replicas, one has to first calculate
\begin{equation}
	\Tr\rho^2=\sum_{\phi_i,\phi_j}\left[\rho\right]_{\phi_i\phi_j}\left[\rho\right]_{\phi_j\phi_i},
\end{equation}
with $\phi_i$ and $\phi_j$ (collective indices) specifying the states of the evaporating black hole. We would omit the unnecessary $\phi$'s and make use of $i,j,\ldots$ instead of $\phi_i,\phi_j,\ldots$ to simplify the notation.
\begin{figure}[t]
	\centering
		\includegraphics[width=\linewidth]{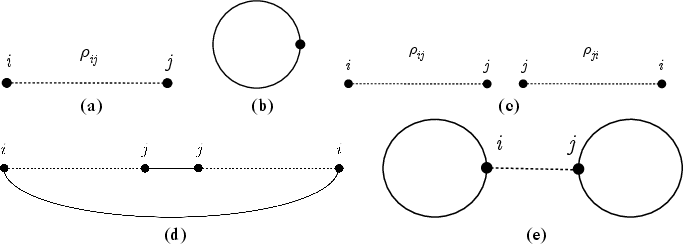}
	\caption{\label{f__rhos1} (a) A visualization of $\rho_{ij}$, which can be regarded as a bridge connecting state $i$ to state $j$. (b) $\Tr(\rho)$ can be obtained from $\rho_{ij}$ by first identifying $i$ and $j$, and then performing a summation over $i=j$. We show it simply by a circle. (c) The product $\rho_{ij}\rho_{ji}$ (no-summation) is shown as the two successive bridges accordingly. (d) Hawking saddle forms one loop and is dominated at initial stages of the evaporation process. (e) Replica wormhole forms two loops by gluing the black holes interiors, and is dominated at final stages of the evaporation process.}
\end{figure}
\par One would think of $\rho_{ij}$ as a bridge connecting state $i$ of the black hole to state $j$, which is schematically shown in Fig.~\ref{f__rhos1}(a). For $\Tr(\rho)$, one should first identify $j$ with $i$ and then perform a summation over all the states, $i$'s. We show it simply by a circle as in Fig.~\ref{f__rhos1}(b). The product $\rho_{ij}\rho_{ji}$ (with no sum over $i$ and $j$) might be regarded as: a bridge connecting state $i$ to $j$, followed by another bridge connecting state $j$ to $i$, which is shown in Fig.~\ref{f__rhos1}(c). However, for $\Tr(\rho^2)$, we should identify the same states ($i\leftrightarrow i$ and $j\leftrightarrow j$) and hence perform summations over $i$ and $j$. There are two different ways of doing the identifications and performing the summations, as shown in Figs.~\ref{f__rhos1}(d) and \ref{f__rhos1}(e):
\begin{enumerate}
	\item Starting from state $i$, moving to state $j$ through the bridge $\rho_{ij}$, identifying $j\leftrightarrow j$, moving back to state $i$ through the bridge $\rho_{ji}$, and finally identifying $i\leftrightarrow i$;
	\item Starting from state $i$, identifying $i\leftrightarrow i$, moving to state $j$ through the bridge $\rho_{ij}$, identifying $j\leftrightarrow j$, and finally moving back to state $i$ through the bridge $\rho_{ji}$.
\end{enumerate}
\par Figs.~\ref{f__rhos1}(d) and \ref{f__rhos1}(e) are topologically different. The first one forms one loop and reproduces Hawking's calculations, called ``Hawking saddle''. However, the second one forms two loops and hence corresponds to some different paths in the context of path integral. It is called ``replica wormhole'', because the interiors of the black holes are joined together forming a kind of wormhole. One would deduce from its topology that it is going to give $(\Tr\rho)^2$ (which is unity if the wave function is truly normalized). As the quantity $\Tr(\rho^2)/(\Tr\rho)^2$ is a measure of ``purity'', Fig.~\ref{f__rhos1}(e) leads to a pure final state in the evaporation process. It should be noticed that the Hawking saddle has an exponentially small contribution to the purity in the final stages. Hence it is not going to get the arguments into trouble. Moreover, there are more replicas present in the calculations, $n=3,\ldots$. Although they contribute to the calculations, they do not affect our qualitative arguments about unitarity of the evaporation process (beneficial discussions are presented in \cite{Kibe2022}). As a matter of fact, the entanglement entropy is mostly given by the Hawking saddle at the initial stages and by the replica wormhole at the final ones, implying the evaporation process to be unitary. Here we have only stated some results of the calculations which would be found in detail in some works such as \cite{Penington2022a,Almheiri2020}.
\par
\begin{figure}[t]
	\centering
		\includegraphics[width=0.47\linewidth]{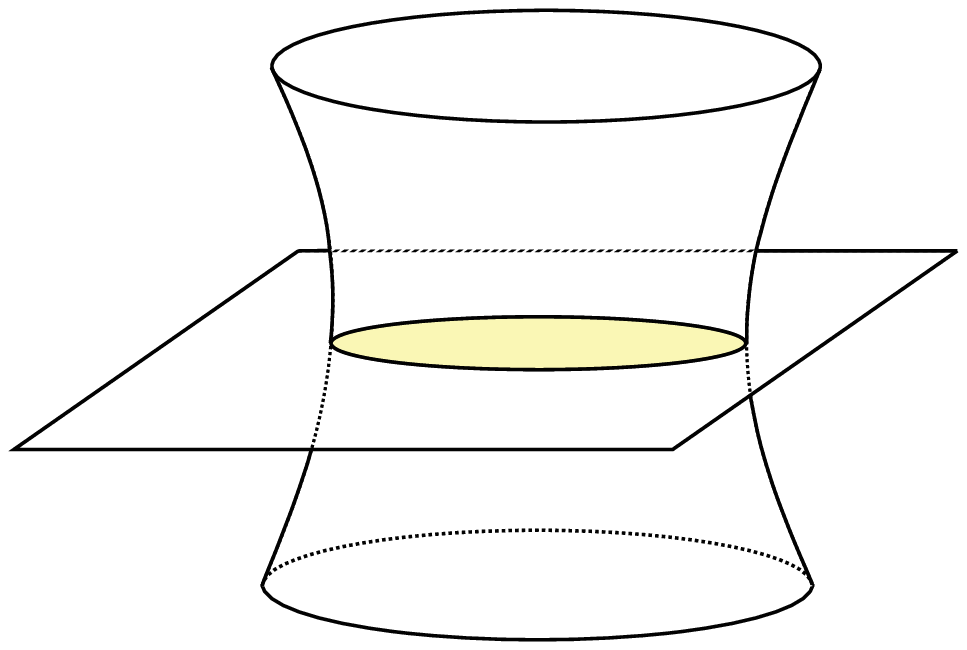}
	\caption{\label{f__RepCut} Replica wormhole glues the black holes interiors. So, a cut is needed when taking the limit $n\rightarrow1$, which is shown by the cutting surface.}
\end{figure}

 We made use of purity instead of entanglement entropy to discuss the unitarity of the evaporation process. The same results will be obtained if one makes use of entanglement entropy. However, it is a bit tricky because of the required analytic continuation of $n$. As the replica wormhole is joining the black holes interiors together, we face with a difficulty in taking the limit $n\rightarrow1$. In fact, a cut is needed in going from $n=2$ to $n=1$, as is shown in Fig.~\ref{f__RepCut}. Such a cut cannot be regarded as a fine operation. This fact would explain the reason for using replica trick: The replica wormhole might be difficult to see in the case of $n$ exactly 1, hence one would miss it as Hawking did!
\par It would be fruitful to summarize this subsection: Replica trick is a mathematical tool which can be used to calculate entanglement entropy of a system. For the special case of an evaporating black hole, this tool will help us to consider all the paths contributing to the respective path integral. To be more specific, the particles that are leaving the black hole have some entanglement with it. This can be regarded as the first order term in the calculation of entanglement entropy, called Hawking saddle. The particles also have some entanglements with those that have left the black hole before. The latter can be seen if one considers higher-order terms, or equivalently, $n=2,3,\ldots$ replicas. One should at least consider $n=2$ replicas to realize that the evaporation process preserves unitarity.
\section{Shape dependence of entanglement entropy}\label{Sec__Shape_Dependence}
Entanglement entropy plays a key role in understanding quantum field theories and quantum gravity \cite{Nomura2018,Faulkner2022}. It represents the level of entanglement between a system and its environment, and depends on the surface that defines the system---the system's boundary. Modifying the boundary has the effect of altering the enclosed fields within the system, inevitably affecting the level of entanglement between the modified system and its new environment. Therefore, it becomes relevant to consider the variations of the entanglement entropy, including its first and higher derivatives, in response to changes in the boundary. Specifically its second variation, called entanglement density, is very interesting due to its relationship with energy density and is studied in some special cases \cite{Leichenauer2018,Faulkner2016,Lewkowycz2018,Rosenhaus2015}. In this section, we derive some consistency requirements for entanglement entropy variations with respect to some boundary deformations. Such consistency requirements help us to obtain relations, among which some would be very fruitful in obtaining the intended Page curve. As it is discussed in the following subsection, the relations might be regarded as some mathematical identities resulting from considering the entanglement entropy as a well-behaved functional of the boundary.
\subsection{Consistency constraints on entanglement entropy variation}\label{sec-consistency}
In this subsection, we obtain relations among different orders of entanglement entropy variation with respect to some boundary deformation. To begin with, we consider a system specified by its boundary, $\chi^\mu$. The system has some entanglement entropy, $S$, which depends on the boundary of the system, $\chi$, through the enclosed matter fields. When the boundary deforms by $\delta\chi$, the enclosed matter fields and the entanglement entropy change accordingly, represented by $\delta S(\chi,\delta\chi)$. If the entanglement entropy is considered to be a well-behaved, smooth functional of the boundary, one would expand it in terms of the second argument, $\delta\chi$, as follows
\begin{equation}\label{deltaS}
		\delta S(\chi,\delta\chi)=\sum_{i=1}^{\infty}\int\delta\chi_{\mu_1}(x_1)\delta\chi_{\mu_2}(x_2) \cdots\delta\chi_{\mu_i}(x_i) I_i^{\mu_1\mu_2\cdots\mu_i}\left[\chi\right](x_1,x_2,\ldots, x_i).
\end{equation}
The coefficients of expansion, $I$'s, depend on the initial boundary $\chi$ (written in brackets) and the points (written in parentheses) through the matter fields. The boundary deformation leads to addition/subtraction of points to/from the system. Moreover, every set of $n$ points has an entanglement entropy and hence contribute to $\delta S$. So we included every order of the boundary deformation---$(\delta\chi)^n,n=1,2,\ldots$---in the above relation through the summation. However, the zeroth order term is omitted due to the fact that there is no entanglement entropy variation without boundary deformation. In addition, the integrations are taken over the whole boundary, $\chi$, so that all the points are taken into account. Contractions are also assumed over the repeated indices, $\mu$'s. Hence the deformations are considered in all the directions.
\par Nothing should change if we first send $\chi$ to $\chi+\delta\chi$ and then send it back to its initial form $\chi$, i.e.,
\begin{equation}\label{CR1}
	\delta S(\chi,\delta\chi)+\delta S(\chi+\delta\chi,-\delta\chi)=0.
\end{equation}
The identity \eqref{CR1} is a consistency condition which should be satisfied for the entanglement entropy to be well-defined. Otherwise, $S(\chi)$ might be multi-valued for a system specified by a unique boundary $\chi$. The consistency condition, Eq.~\eqref{CR1}, can be transformed into relations among the integrands of Eq.~\eqref{deltaS}, $I$'s. To do so, we first write expansions for $\delta S(\chi,\delta\chi)$ and $\delta S(\chi+\delta\chi,-\delta\chi)$ as of Eq.~\eqref{deltaS}
\begin{equation}
	\begin{split}
			&\delta S(\chi,\delta\chi)= \sum_{i=1}^{\infty}\int\delta\chi_{\mu_1}(x_1)\cdots\delta\chi_{\mu_i}(x_i) I_i^{\mu_1\cdots\mu_i}\left[\chi\right](x_1,\ldots, x_i), \\
			&\delta S(\chi+\delta\chi,-\delta\chi) = \sum_{i=1}^{\infty}\left(-1\right)^{i}\int\delta\chi_{\mu_1}(x_1)\cdots\delta\chi_{\mu_i}(x_i) I_i^{\mu_1\cdots\mu_i}\left[\chi+\delta\chi\right](x_1,\ldots, x_i).
	\end{split}
\end{equation}
We assume that $I[\chi+\delta\chi]$ is a well-behaved functional of $\chi$, such that we can write the following expansion for it
\begin{equation}
	\begin{split}
		I\left[\chi+\delta\chi\right]= \sum_{j=0}^{\infty}\int\frac{1}{j!}&\frac{\partial^j I}{\partial\chi_{\mu_1}(x_1)\partial\chi_{\mu_2}(x_2)\cdots\partial\chi_{\mu_j}(x_j)}\left[\chi\right]\\
		&\cdot \delta\chi_{\mu_1}(x_1)\delta\chi_{\mu_2}(x_2)\cdots\delta\chi_{\mu_j}(x_j).
	\end{split}
\end{equation}
The derivatives in the above expansion are functional derivatives and people mostly use $\delta$ for them. However, we made use of $\partial$ to prevent confusion with the $\delta$'s used for entropy variations and boundary deformations. Finally we request the identity to be held independently for each order of boundary deformation and in each set of points due to the arbitrariness of deformation, $\delta\chi(x)$. So we collect coefficients of $\delta\chi_{\mu_1}(x_1)\cdots\delta\chi_{\mu_m}(x_m)$ to obtain the following set of relations
\begin{equation}\label{MasterEq}
		\left(1+(-1)^m\right)I_m^{\mu_1\cdots\mu_m}(x_1,\ldots,x_m)=\sum_{l=1}^{m-1}\frac{(-1)^{l+1}}{(m-l)!}\frac{\partial^{m-l}I_l^{\mu_1\cdots\mu_l}(x_1,\ldots,x_l)}{\partial\chi_{\mu_{l+1}}(x_{l+1})\cdots\partial\chi_{\mu_m}(x_m)},
\end{equation}
for $m=1,2,\ldots$. A relation that we will mostly use in this work can be obtained from Eq.~\eqref{MasterEq} by setting $m=2$,
\begin{equation}\label{Eq-I1I2}
	2I_2^{\mu\nu}(x_1,x_2)=\frac{\partial I_1^\mu(x_1)}{\partial\chi_\nu(x_2)}.
\end{equation}
From now on, we drop the $\mu_i$ indices and the $x_i$ dependencies for briefness.
\par The left hand side of Eq.~\eqref{MasterEq} vanishes for odd values of $m$. Hence, it does not express any odd-order $I$ in terms of the derivative terms. Nevertheless, such expressions would be found by using some other consistency requirement: Let first deform the boundary $\chi$ to $\chi+\delta\chi$, then send $\chi+\delta\chi$ to $\chi+2\delta\chi$, and finally turn it back to its initial form, $\chi$. As before, one would expect that nothing should change under these three successive transformations, i.e.,
\begin{equation}\label{CR2}
	\delta S(\chi,\delta\chi)+\delta S(\chi+\delta\chi,\delta\chi)+\delta S(\chi+2\delta\chi,-2\delta\chi)=0.
\end{equation}
This consistency condition leads to relations, among which some might not be independent of those of Eq.~\eqref{MasterEq} (for example, it gives Eq.~\eqref{Eq-I1I2} again). Desirably, some expressions would be obtained for some odd-order $I$'s in terms of the derivative terms, such as
\begin{equation}
	6I_3=-\frac{7}{2}\cdot\frac{\partial^2 I_1}{\partial \chi^2}+9\cdot\frac{\partial I_2}{\partial \chi},
\end{equation}
which can also be combined with the $m=3$ case of Eq.~\eqref{MasterEq} to obtain
\begin{equation}\label{Eq__I3I2I1}
	6I_3=2\cdot \frac{\partial I_2}{\partial\chi}=\frac{\partial^2I_1}{\partial\chi^2}.
\end{equation}
\par Following Eqs.~\eqref{CR1} and \eqref{CR2}, more consistency identities can be written for entanglement entropy variations with respect to some boundary deformations. In general, one would expect
\begin{equation}\label{Eq__consistency}
		\delta S(\chi,\delta\chi)+\delta S(\chi+\delta\chi,\delta\chi)+\cdots +\delta S(\chi+(n-1)\delta\chi,\delta\chi)+\delta S(\chi+n\delta\chi,-n\delta\chi) =0,
\end{equation}
for $n=1,2,\ldots$. Again, one can transform the consistency identities into relations among $I$'s, following the steps made in the case of $n=1$. As stated in the case of $n=2$, not all the relations obtained in this way are independent---Some of them would be reproduced using some others.
\subsection{An interpretation for \texorpdfstring{$I_2$}{I\_2}}\label{sec__I_2}
\begin{figure}[t]
	\centering
	\includegraphics[width=.6\linewidth]{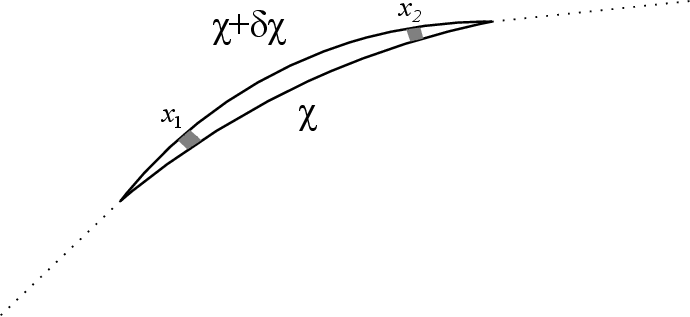}
	\caption{\label{f-Chi} A boundary deformation of a system is shown. The two segments specified by gray squares have a mutual entanglement entropy, which is not going to change the entanglement entropy of the system. This mutual entanglement, however, is considered two times through the first order terms, $I_1$'s. The $I_2$ terms, inevitably, should cancel these over-calculations.}
\end{figure}
Let us focus on Fig.~\ref{f-Chi} to find an interpretation for $I_2$ in Eq.~\eqref{deltaS}. The boundary is changed from $\chi$ to $\chi+\delta\chi$ so that the system is enlarged. The specified segments (shown by gray squares) participate in $\delta S$ through
\begin{equation}\label{eq__deltaS12}
	\begin{split}
			\left.\delta S\right|_{x_1x_2}=\;&I_1^\mu(x_1)\delta\chi_\mu(x_1)+I_1^\mu(x_2)\delta\chi_\mu(x_2)\\
			&+I_2^{\mu\nu}(x_1,x_2)\delta\chi_\mu(x_1)\delta\chi_\nu(x_2)+I_2^{\mu\nu}(x_2,x_1)\delta\chi_\mu(x_2)\delta\chi_\nu(x_1)+\cdots.
	\end{split}
\end{equation}
The $I_1$ terms are standing to take the entanglement entropy variation at the first order. So, they are adding the entanglement entropy of each segment with outside (environment), while subtracting the entanglement entropy of that segment with inside (system). However, each couple of segments added to the system has a mutual entanglement entropy which is not going to change the entanglement entropy of the system, but is added to $\delta S$ for two times through $I_1(x_1)$ and $I_1(x_2)$. This mutual entanglement entropy is proportional to both $\delta\chi(x_1)$ and $\delta\chi(x_2)$, hence it is proportional to the product $\delta\chi(x_1)\delta\chi(x_2)$. Therefore, a modification should occur in the second order to cancel these over-calculations. In other words, the $I_2$ terms are standing to cancel the added mutual entanglement entropies, i.e.,
\begin{equation}
	\begin{split}
		I_2^{\mu\nu}(x_1,x_2)+I_2^{\nu\mu}(x_2,x_1))&\delta\chi_\mu(x_1)\delta\chi_\nu(x_2)=
		\\
		&-2 (\text{mutual entangl. ent. of segments 1 \& 2}).
	\end{split}
\end{equation}
The expression at the left-hand side of the above equation can be written in a more brief manner using the following definition 
\begin{equation}\label{Eq__Sym}
		\text{Sym}\{I_2^{\mu\nu}(x_1,x_2)\}:=\frac{1}{2}\left(I_2^{\mu\nu}(x_1,x_2)+I_2^{\nu\mu}(x_2,x_1)\right).
\end{equation}
Thus, we can simply write
\begin{equation}\label{Eq__I_2__Interpretation}
		\mathrm{Sym}\{I_2^{\mu\nu}(x_1,x_2)\}\delta\chi_\mu(x_1)\delta\chi_\nu(x_2)
		 =-\left(\text{mutual entangl. ent. of segments 1 \& 2}\right),
\end{equation}
which is the desired interpretation for $I_2$. One can combine Eqs.~\eqref{Eq-I1I2} and \eqref{Eq__I_2__Interpretation} to obtain
\begin{align}
		\text{mutual entangl. ent. of segments 1 \& 2} &= \nonumber\\
		 -\frac{1}{4}\left(\frac{\partial I_1^\mu(x_1)}{\partial\chi_\nu(x_2)} + \frac{\partial I_1^\nu(x_2)}{\partial\chi_\mu(x_1)}\right)&\delta\chi_\mu(x_1)\delta\chi_\nu(x_2),
\end{align}
which gives an interpretation for the combination in the parentheses.
\subsection{Interpreting the higher-order terms, \texorpdfstring{$I_3$}{I\_3} etc}\label{Subsec__Interpret_In}
To complete the discussion, we want to note that there are more over-calculations present in the first order terms, to be canceled with including higher-order terms, $I_3$ etc. To clarify, let consider three segments added to the system simultaneously, and write $\delta S|{x_1x_2x_3}$, like what we did in Eq.~\eqref{eq__deltaS12}. This time we write the terms up to the third order explicitly, i.e.,
\begin{equation}
	\begin{split}
		\delta &S|_{x_1x_2x_3}=I_1(x_1)\delta\chi(x_1)+I_1(x_2)\delta\chi(x_2)+I_1(x_3)\delta\chi(x_3)\\
		&+2I_2(x_1,x_2)\delta\chi(x_1)\delta\chi(x_2)+2I_2(x_2,x_3)\delta\chi(x_2)\delta\chi(x_3)\\
		&+2I_2(x_1,x_3)\delta\chi(x_1)\delta\chi(x_3)+(3!)I_3(x_1,x_2,x_3)\delta\chi(x_1)\delta\chi(x_2)\delta\chi(x_3)+\cdots,
	\end{split}
\end{equation}
where the coefficients $I_2$ and $I_3$ are assumed to be symmetric with respect to their arguments, like what we did in Eq.~\eqref{Eq__Sym}. It is possible to have a non-zero tripartite entanglement among the segments. In such a situation, one is adding this tripartite entanglement entropy three times through $I_1$'s in the first line of the above equation. The $I_3$ terms should cancel these over-calculations, i.e.,
\begin{equation}
	\begin{split}	\left(3!\right)I_3(x_1,x_2,x_3)\delta\chi(x_1)&\delta\chi(x_2)\delta\chi(x_3)=\\
		-3&\left(\text{tripartite entangl. ent. of segments 1,2 and 3}\right).
	\end{split}
\end{equation}
Continuing this procedure, one can deduce in general that
\begin{equation}
	\begin{split}
	\text{Sym}\{I_n(x_1,x_2,\ldots,x_n)\}&\delta\chi(x_1)\delta\chi(x_2)\cdots\delta\chi(x_n) =\\
	-\frac{n}{n!}&\left(n\text{-partite entangl. ent. of segments }1,2,\ldots,n\right),
	\end{split}
\end{equation}
which is a generalization of Eq.~\eqref{Eq__I_2__Interpretation} for $n=2,\ldots$.
\par Summarizing subsections \ref{sec-consistency}-\ref{Subsec__Interpret_In}: The different orders of the entanglement entropy variation with respect to some boundary deformation in Eq.~\eqref{deltaS}, $I_n\delta\chi^n$'s, are giving the $n$-partite entanglement entropies among the segments that are specified by the $\delta\chi$'s. At the other hand, there are some consistency requirements, Eq.~\eqref{Eq__consistency}, that result in some relations among $I_n$'s and $\partial^m I_{n-m}/\partial \chi^m$'s for $m=1, 2, \cdots n-1$. So performing $m$ derivatives on $(n-m)$-partite entanglement entropy with respect to the boundary would raise its order of entanglement from $n-m$ to $n$, i.e.,
\begin{equation}
	n\text{-partite entangl. ent.} \quad\sim\quad \frac{\partial^m}{\partial\chi^m}\left[	(n-m)\text{-partite entangl. ent.}\right].
\end{equation}
Hence the different orders of entanglement are related to each other through some consistency requirements.
\section{Entanglement entropy of a radiating black hole and the Page curve}\label{Replica__EE}
In this section, our goal is to derive an expression for the entanglement entropy of a (3+1)-dimensional evaporating Schwarzschild black hole under the smoothness assumption. We utilize the replica trick and make the following assumptions: (1) In the replicated system, the entanglement entropy of each black hole scales with its area; (2) The black hole's entanglement entropy will not exceed the Bekenstein-Hawking entropy. The \nth{1} holographic assumption can be understood through the {\sl Central Dogma} in black hole physics, which posits that: As seen from the outside, a black hole can be described in terms of a quantum system with $Area/4G$ degrees of freedom, which evolves unitarily under time evolution \cite{Almheiri2021}. On the other hand, the \nth{2} holographic assumption stems from the principle that fine-grained or entanglement entropy is always less than or equal to coarse-grained or thermodynamic entropy.
\par
\begin{figure}[t]
	\centering
	\includegraphics[width=\linewidth]{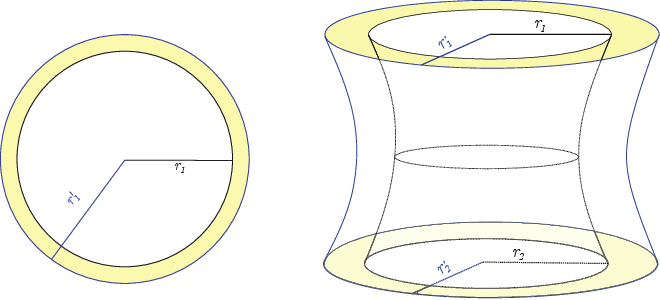}
	\caption{\label{f-Replica} Left: A (2+1)-dimensional disk is visualized in two spatial dimensions, representing a Stereographic transformation of the event horizon of a black hole. The yellow-shaded region is added to the disk by a radius variation, which leads to some entanglement entropy variation. Right: $n=2$ replicas are shown, although with different radii! The yellow-shaded regions represent radii variations which lead to some variation of the entanglement entropy of the whole system. The connecting curves (wormhole) represent some entanglement between the disks in the sprite of ER=EPR.}
\end{figure}

 To enhance the visual representation, we can transform the spherical event horizon of each black hole into a double-sided disk. This approach involves using two Stereographic maps—one for the top hemisphere and another for the bottom hemisphere. The resulting disk has a radius proportional to the black hole radius, with the constant of proportionality absorbed into other constants. Importantly, the following arguments remain valid independent of the transformation. Even without the Stereographic projection, one can follow the arguments and end in the results obtained for the entanglement entropy of the black hole\footnote{The exclusive purpose of the transformation is to illustrate the replicated system, as shown in Fig.~\ref{f-Replica}.}.
\par In the light of our $\nth{1}$ holographic assumption, we consider a disk of radius $r_1$ in two spatial dimensions and call it $\mathcal{A}$\footnote{We first analyze one disk and then replicate it to be in accord with the replicated black hole system.}. The entanglement entropy of $\mathcal{A}$ depends on its radius, $r_1$, as well as on the details of the field theory that describes it. As discussed in Sec.~\ref{Sec__Intro}, we are making use of the holographic assumptions alongside the consistency relations of Sec.~\ref{Sec__Shape_Dependence}, to obtain a smooth curve for the entanglement entropy of the black hole. The consistency conditions enforce the entropy to be well-behaved in the middle points, because by expanding $S$ in Sec.~\ref{Sec__Shape_Dependence}, we are considering it to be smooth. Some higher-order modifications would occur, one of which is explained in subsection \ref{Sec__higher_replicas}.
\par Let start the procedure by variating the radius of the disk $\mathcal{A}$, from $r_1$ to $r_1^\prime=r_1+\delta r_1$, Fig.~\ref{f-Replica}, left. The variation of the entanglement entropy takes the form
\begin{equation}
	\delta S_\mathcal{A}=I_1(r_1)\delta r_1,
\end{equation}
where $I_1$ contains the unspecified details of the field theory. The points added to $\mathcal{A}$, shown by yellow in the figure, have some entanglement with both $\mathcal{A}$ and the environment. Thus $I_1(r_1)\delta r_1$ contains the entanglement entropies of the points with the environment and $\mathcal{A}$, with positive and negative signs respectively. Hence it should be proportional to the number of the added points, or equivalently, to the added area, i.e.,
\begin{equation}\label{eq__disk1}
	I_1(r_1)\delta r_1=c r_1\delta r_1,
\end{equation}
where we absorbed a factor of $2\pi$ in $c$. Moreover, $c$ contains information about the field theory and, consequently, depends on $r_1$ in general. So we would expand it in terms of $r_1$, or equivalently, make use of the replica trick according to the discussions passed in Sec.~\ref{Sec__whyrt}.
\par Let consider $n=2$ replicas and label them by 1 and 2, as in Fig.~\ref{f-Replica} right. We take the radii of the disks to be different at first; however, we set them equal after we find a proper expression for the entanglement entropy variation, $\delta S^{(n=2)}$. This would be considered as a trick for obtaining some relations much easier. The entanglement entropy variation due to the radii variations takes the general form
\begin{equation}
	\delta S^{(n=2)}=I_1(r_1;r_2)\delta r_1+I_1(r_2;r_1)\delta r_2+2I_2(r_1,r_2)\delta r_1\delta r_2,		
\end{equation}
where $r^\prime_1$ and $r^\prime_2$ in the figure are substituted by $r_1+\delta r_1$ and $r_2+\delta r_2$ respectively. The first (second) term is standing to take the entanglement entropy variation between the disk 1 (2), shown by darker (lighter) yellow, with the environment into account; while the $I_2$ term (symmetric under $r_1\leftrightarrow r_2$) is standing to take the entanglement entropy variation between the disks, 1 and 2, into account. The connecting curves in Fig.~\ref{f-Replica}, right, are referring to such an entanglement in the spirit of ER=EPR. We have separated $r_1$ and $r_2$ dependencies of $I_1(r_1;r_2)$ (or $I_1(r_2;r_1)$) by ``;'' to emphasize that: Although the term $I_1(r_1;r_2)\delta r_1$ is standing for the entanglement change of the disk 1 and the environment, it implicitly depends on the value of $r_2$ due to the presence of some entanglement between the disks, 1 and 2. One can deduce, as in Eq.~\eqref{eq__disk1}, that
\begin{equation}\label{eq__I_1}
	I_1(r_1;r_2)\delta r_1= \frac{\alpha(r_2) r_1}{G}\delta r_1,
\end{equation}
and
\begin{equation}\label{eq__I_2}
	I_1(r_2;r_1)\delta r_2= \frac{\alpha(r_1) r_2}{G}\delta r_2,
\end{equation}
where $G$ is the Newton's constant in (3+1) dimensions, inserted to make $\alpha$ dimensionless\footnote{We used $\alpha$ in both Eqs.~\eqref{eq__I_1} and \eqref{eq__I_2} due to the symmetry $1\leftrightarrow2$.}. As one can see, we dropped the dependence of $\alpha$ on the first argument of $I_1$ according to the discussion made after Eq.~\eqref{eq__disk1}. One can also deduce that the $I_2(r_1,r_2)\delta r_1\delta r_2$ term should be proportional to both the number of the points added to disk 1 and disk 2; or equivalently, it should be proportional to the multiplication of the areas of the two yellow-shaded regions in Fig.~\ref{f-Replica} right, i.e.,
\begin{equation}
	I_2(r_1,r_2)\delta r_1\delta r_2= -\frac{\zeta r_1 r_2}{G^2}\delta r_1\delta r_2,
\end{equation}
where $\zeta$ is a dimensionless constant and positive according to Eq.~\eqref{Eq__I_2__Interpretation}---The minus sign is inserted to make it positive. Making use of Eq.~\eqref{Eq-I1I2}, one can find the following relation between the unknown coefficients, $\alpha$ and $\zeta$
\begin{equation}
	\frac{d \alpha(r)}{d r}=-\frac{2\zeta r}{G},
\end{equation}
in which the functional derivative with respect to the boundary, $\partial/\partial\chi(x)$, is replaced by the common derivative $d/dr$. The reason is that the boundary variation in this special case is symmetric, independent of the angular coordinate: $\delta\chi(x)=\delta r(\theta)=dr$. Integrating the above relation leads to
\begin{equation}
	\alpha(r)= -\frac{\zeta r^2}{G}+\gamma,
\end{equation}
with $\gamma$ being a dimensionless constant. Now we can write down the entropy variation as
\begin{equation}
		\delta S^{(n=2)} = \left(\gamma-\frac{\zeta r_2^2}{G}\right)\frac{r_1}{G}\delta r_1-\frac{2\zeta r_1 r_2}{G^2}\delta r_1\delta r_2 +\left(\gamma-\frac{\zeta r_1^2}{G}\right)\frac{r_2}{G}\delta r_2.
\end{equation}
It is an easy task to find an expression for $\delta S_\mathcal{A}$. To do so, let look at the terms participating in $\delta S^{(n=2)}$. We have two $I_1$ terms which is a direct consequence of having $n=2$ replicas. Moreover, every pair of replicas will lead to a $2I_2$ term. Thus, returning the $n$-dependence and setting $r_1=r_2=r$, we have
\begin{equation}
	\delta S^{(n=2\rightarrow n)}= n\left(\gamma-\frac{\zeta r^2}{G}\right)\frac{r}{G}\delta r-2 \binom{n}{2}\frac{\zeta r^2}{G^2}\delta r^2,
\end{equation}
where the superscript index $(n=2\rightarrow n)$ is indicating that we have obtained the above equation from $n=2$ replicas. There are also other terms participating in $\delta S$ which would be appear when we consider more replicas, $n=3$ etc. Performing the limit $n\rightarrow 1$, the above equation gives
\begin{equation}
	\delta S^{(2)}=\delta S_\mathcal{A}\approx \left(\gamma-\frac{\zeta r^2}{G}\right)\frac{r}{G}\delta r,
\end{equation}
where the superscript $(2)$ refers to the fact that we have considered the terms up to $I_2$.
\par It is worth noting that although the term $I_2(r_1,r_2)$ disappeared from our calculations in the limit $n\rightarrow 1$, it played a crucial role in deriving the above expression for the entanglement entropy variation: The second term within the parentheses underscores the significance of including additional replicas to account for higher-order contributions. The entanglement entropy can easily be obtained from the above equation
\begin{equation}\label{Eq__pme_constants}
	S^{(2)}(r)= \frac{\gamma r^2}{2G}-\frac{\zeta r^4}{4G^2}.
\end{equation}
The two unknown constants $\zeta$ and $\gamma$ would be found by using the following constraints on the black hole: i) The entanglement entropy vanishes at $r=r_0$---the initial radius of the black hole---which means that the black hole was initially in a pure state, i.e.,
\begin{equation}
	S(r=r_0)=0;
\end{equation}
ii) The entanglement entropy will not exceed the Bekenstein-Hawking entropy, $A/4G$---At most, it saturates the bound. Thus we have
\begin{equation}
\text{Max}_{r}\left\{\frac{4GS(r)}{A(r)}\right\}=1.	
\end{equation}
It is obvious from Eq.~\eqref{Eq__pme_constants} that the maximum occurs when $r\rightarrow0$. Applying these two boundary conditions for $r=0$ and $r=r_0$, one obtains $\gamma=2\pi$ and $\zeta=4\pi G/r_0^2$, hence
\begin{equation}\label{Eq__pme}
	S^{(2)}(r)= \frac{\pi r^2}{G}-\frac{\pi r^4}{G r_0^2}=\frac{\pi r^2}{G}\left(1-\frac{r^2}{r_0^2}\right).
\end{equation}
We can easily convert the $r$-dependence of $S$ into the time-dependence. To do so, we would make use of $r=2GM$, with $M$ being mass of the black hole variating by time as given in \cite{Page2013}
\begin{equation}\label{Eq__MtM0}
	M(t)=M_0\left(1-\frac{t}{t_\text{decay}}\right)^{1/3},
\end{equation}
where $M_0$ is the initial mass of the black hole and $t_\text{decay}$ is the time duration of the evaporation process as measured by an asymptotic observer. Hence we find
\begin{equation}\label{Eq__pmet}
		S^{(2)}(t)=\pi G M_0^2\left(1-\frac{t}{t_\text{decay}}\right)^{2/3}\left[1-\left(1-\frac{t}{t_\text{decay}}\right)^{2/3}\right].
\end{equation}
In the limit $r\ll r_0$, or equivalently $t\lesssim t_\text{decay}$, Eqs.~\eqref{Eq__pme} and \eqref{Eq__pmet} give
\begin{equation}\label{Eq__pmea1}
	\begin{split}
		S^{(2)}&\left(r\ll r_0\right)\approx \frac{\pi r^2}{G},\\ S^{(2)}&\left(t\lesssim t_\text{decay}\right) \approx 4\pi G M_0^2\left(1-\frac{t}{t_\text{decay}}\right)^{2/3},
	\end{split}
\end{equation}
the first of which is the Bekenstein-Hawking entropy formula $S=A/4G$. In the other limit of interest $r\lesssim r_0$, or equivalently $t\ll t_\text{decay}$, we have
\begin{equation}\label{Eq__pmea2}
	\begin{split}
		&S^{(2)}\left(r\lesssim r_0\right)\approx \frac{2\pi}{G}r_0\left(r_0-r\right),\\
		 &S^{(2)}\left(t\ll t_\text{decay}\right)\approx \frac{8\pi GM_0^2}{3}\frac{t}{t_\text{decay}}.	
	\end{split}
\end{equation}
\par One can compare the obtained results with those of Page \cite{Page2013}
\begin{align}
	S_\text{P}(t)= 4\pi G\beta M_0^2&\left[1-\left(1-\frac{t}{t_\text{decay}}\right)^{2/3}\right]\theta\left(t_\text{P}-t\right)\nonumber\\
	&\qquad+4\pi G M_0^2\left(1-\frac{t}{t_\text{decay}}\right)^{2/3}\theta\left(t-t_\text{P}\right), \label{Eq__Paget}
\end{align}
where $\theta$ is the Heaviside step function, $\beta\approx 1.48472$ and
\begin{equation}
	t_\text{P}=\left[1-\left(\frac{\beta}{\beta+1}\right)^{3/2}\right] t_\text{decay}.
\end{equation}
Converting the time-dependence into $r$-dependence via Eq.~\eqref{Eq__MtM0} alongside $M=r/2G$, one obtains
\begin{equation}\label{Eq__Pager}
		S_\text{P}(r)= \frac{\pi\beta}{G} \left(r_0^2-r^2\right)\theta\left(r-r_\text{P}\right)+\frac{\pi}{G} r^2\theta\left(r_\text{P}-r\right),
\end{equation}
where
\begin{equation}
	r_\text{P}=\left(1-\frac{t_\text{P}}{t_\text{decay}}\right)^{1/3}r_0.
\end{equation}
\begin{figure}[t]
	\centering
	\includegraphics[width=\linewidth]{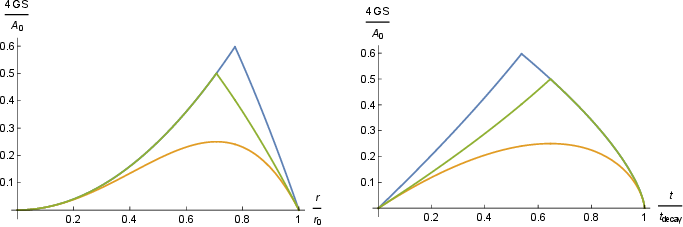}
	\caption{\label{f-Pcs} Comparing the obtained results from $n=2$ replicas with those of Page: The orange curves are the results from $n=2$ replicas (Eqs.~\eqref{Eq__pme} and \eqref{Eq__pmet}); while the blue curves are those of Page (Eqs.~\eqref{Eq__Paget} and \eqref{Eq__Pager}). The green curves are the two limiting behaviors of the orange curves (Eqs.~\eqref{Eq__pmea1} and \eqref{Eq__pmea2}) extrapolated to the inner region to intersect each other.}
\end{figure}\\
The obtained results, Eqs.~\eqref{Eq__pme} and \eqref{Eq__pmet}, along with the results of Page, Eqs.~\eqref{Eq__Paget} and \eqref{Eq__Pager}, are depicted in Fig.~\ref{f-Pcs} as orange and blue curves, respectively. As discussed in Sec.~\ref{Sec__Intro}, the Page curve is derived by considering the thermodynamic entropies of the radiation and the black hole and transitioning between them, as illustrated in Fig.~\ref{f__Page}. Hence, we have extrapolated the two limiting behaviors (given by Eqs.~\eqref{Eq__pmea1} and \eqref{Eq__pmea2}) of Eqs.~\eqref{Eq__pme} and \eqref{Eq__pmet} using green lines for comparison with the results of Page. As observed, they exhibit similarity for low to moderate values of $r$, or correspondingly, for medium to high values of time. Deviations can be observed in the intermediate regions of the plots. However, when dealing with $n=3$ replicas, a modification arises whereby the curves (blue and green) can potentially coincide. This potential will be thoroughly examined in subsection \ref{Sec__higher_replicas}, leading to the depiction of the curves illustrated in Fig~\ref{f-Pcs--Modified}.
\par As is discussed in Sec.~\ref{Sec__Intro}, the Page curve has a singular (breaking) point at the Page time arising from the phase transition between a disconnected Euclidean topology and a connected one. We treated $S$ as a well-behaved function, such that our results inhibit any singular point. Nevertheless, we benefited from Page's assumptions, especially for the limiting behaviors. Consequently, the obtained curve, shown in green in Fig.~\ref{f-Pcs--Modified}, can be regarded as a smooth version of the Page curve, corresponding to a gradual transition.
\begin{figure}[t]
	\centering
	\includegraphics[width=\linewidth]{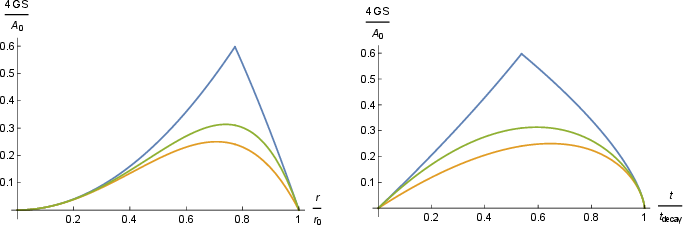}
	\caption{\label{f-Pcs--Modified} Comparing the obtained results from $n=2$ and $n=3$ replicas with those of Page: The orange curves are the results from $n=2$ replicas (Eqs.~\eqref{Eq__pme} and \eqref{Eq__pmet}); while the green curves are those from $n=3$ replicas (Eqs.~\eqref{Eq__pme2r} and \eqref{Eq__pme2t}) and the blue curves are the results of Page (Eqs.~\eqref{Eq__Paget} and \eqref{Eq__Pager}). The limiting behaviors of the orange curves are shown in Fig.~\ref{f-Pcs}, while the limiting behaviors of the green curves coincide with the results of Page.}
\end{figure}
\par One can define a time in our results that corresponds to the Page time, denoted as $t_*$. As there are no breaking points, it can be identified as the time when the entanglement entropy reaches its maximum value. The values of $t_*$ for both cases of considering $n=2$ and $n=3$ replicas are reported in Table \ref{Table}.
\par According to the Page curve, quantum state of the radiation is purely thermal prior to the Page time. However, as our curves (orange and green) in Fig.~\ref{f-Pcs--Modified} lie below the Page curves (blue), one would conclude that the radiation has more informational content when the transition is gradual. Some useful discussions are presented in \cite{Kudler-Flam2022} about the informational content of a black hole's radiation. The difference of the maximum entropy values for the Page curve and for the obtained curves, reported in Table \ref{Table}, can be regarded as a measure of such additional information. Notably, when the transition is gradual, the informational content of the emitted radiation increases by approximately 50\% at the Page time. To be more concrete, according to the Page curve, the evaporation process is quite random prior to the Page time which leads to more and more loss of information about the precise state of the remaining black hole. However, the information begins to come out after the Page time, which prevents the evaporation process to be much random. In other words, the randomness of the evaporation process drops strictly when passing the Page time. The situation is quite different for the smooth curves---Fig~\ref{f-Pcs--Modified}, green. In that case, the evaporation process leads to a mixture of both losing and gaining information about the remaining black hole in each step. So, in the case of a gradual transition, the randomness of the evaporation process does not undergo a sharp change when passing the maximum point. This aligns with our previous discussion in Sec.~\ref{Sec__Intro} and the proposal that different parts of the system's state may have undergone the phase transition or not \cite{Akers2021}.
\begin{table}[ht!]
\tbl{The location and the height of maximum points are stated in curves $S_\text{P}(t)$, $S^{(2)}(t)$ and $S^{(3)}(t)$. The quantity $(S_\text{P}(t)-S(t))/S_\text{P}(t)$, for any time prior to the Page time, can be regarded as the amount of the additional information about the quantum state of the black hole that is stored in the radiation. This quantity is also specified for the curves at their respective maxima, which occur at different times.}{
	\begin{tabular}{@{}cccc@{}}\toprule
		& Value ($\#/t_\text{decay}$) & Entropy ($S/4\pi GM_0^2$) & $(S_\text{P}(t_\text{P})-S)/S_\text{P}(t_\text{P})$ \\ \colrule
		$t_\text{P}$ & $0.54$ & $0.60$ & 0 \\
		$t_*^{(2)}$ & $0.65$ & $0.25$ & $0.58$ \\
		$t_*^{(3)}$ & $0.59$ & $0.31$ & $0.48$ \\ \botrule
	\end{tabular}\label{Table}}
\end{table}
\par The Page curve can be understood as a consequence of Page's theorem. This theorem states that in a bipartite system described by the Hilbert space $H_{AB} = H_A \otimes H_B$, a randomly chosen pure state in $H_{AB}$ is likely to be very close to maximally entangled when the dimension of one subsystem is significantly larger than that of the other \cite{Page1993a,Harlow2016}. As a result, the Page curve effectively characterizes the evolution of an evaporating black hole when the Hilbert space of either the radiation or the remaining black hole is significantly smaller than that of the other component. Specifically, the curve holds when the entanglement entropy remains approximately less than or equal to $\mathcal{O}(10^{-1})$ of $A_0/4G$\footnote{At this point, the logarithm of the dimension of the Hilbert space associated with the radiation or the remaining black hole constitutes 10\% of the entire system.}. In the range between these segments, the curve receives a correction of order $1$ \cite{Harlow2016,Akers2021}. It is important to recognize that the correction is expected to diminish the curve, consistent with the principle that fine-grained entropy should always be less than or equal to coarse-grained entropy. Notably, the green curve depicted in Fig.~\ref{f-Pcs--Modified} aligns well with these projections. It accurately represents the designated points and behaves as anticipated in the intermediate segments.
%%%
\subsection{Considering \texorpdfstring{$n=3$}{n=3} replicas}\label{Sec__higher_replicas}
%%%%
As mentioned earlier in Sec.~\ref{Sec__Replica} and this section, considering more replicas ($I_3$, etc.) affect our calculations by some higher-order terms ($\mathcal{O}(r^6)$). As the first correction to Eq.~\eqref{Eq__pme}, let consider $n=3$ replicas. In this situation a term $I_3$ corresponding to some tripartite entanglement should be included, which through the same arguments as of $I_1$ and $I_2$ should be of the form
\begin{equation}
	I_3(r_1,r_2,r_3)\delta r_1\delta r_2\delta r_3=-\frac{\eta r_1 r_2 r_3}{6G^3}\delta r_1\delta r_2\delta r_3,
\end{equation}
with $\eta$ being a positive dimensionless constant. Such a term will give an $r^6$ correction to Eq.~\eqref{Eq__pme_constants}. Hence the modified equation becomes
\begin{equation}\label{Eq__S3}
		S^{(3)}(r)= \frac{\gamma^{(3)} r^2}{2G}-\frac{\zeta^{(3)} r^4}{4G^2}-\frac{\eta r^6}{6G^3},
\end{equation}
where the superscript $(3)$ refers to the fact that we have considered the terms up to $I_3$.
\par One would expect the constant $\eta$ to be computed directly from the gravitation theory\footnote{The direct computation of $\eta$ needs specifying the gravitation theory. Here we just made use of some holographic assumptions instead. It also might need some heavy and numerical computations.}. However, we want just to show that it is possible for such a term to modify our results to be in agreement with those of Page for $r\lesssim r_0$, or equivalently for $t\ll t_\text{decay}$. The three constants of Eq.~\eqref{Eq__S3} should be determined by three boundary conditions. We have two of them (as before) and assume that the third one is enforcing the $r\lesssim r_0$ limiting behavior of the entanglement entropy to be matched with the Page's result. Hence the boundary conditions are
\begin{equation}\label{Eq__S_BC}
		\text{Max}_{r}\left\{\frac{4GS(r)}{A(r)}\right\}=1,\qquad S(r=r_0)=0,\qquad
		\left.\frac{dS(r)}{dr}\right|_{r\rightarrow r_0}=-\frac{2\pi \beta}{G}r_0,
\end{equation}
which result in
\begin{equation}\label{Eq__new-consts}
	\gamma^{(3)}=2\pi =\gamma,\qquad \zeta^{(3)}=\frac{4\pi G(2-\beta)}{r_0^2},\qquad \eta=\frac{6\pi G^2(\beta-1)}{r_0^4},
\end{equation}
that give
\begin{equation}\label{Eq__pme2r}
	S^{(3)}(r)=\frac{\pi r^2}{G}-(2-\beta)\frac{\pi r^4}{Gr_0^2}-(\beta-1)\frac{\pi r^6}{Gr_0^4}.
\end{equation}
Replacing from Eq.~\eqref{Eq__MtM0}, we find the following expression for $S^{(3)}(t)$
\begin{align}
		S^{(3)}(t)=4\pi G M_0^2\left[\left(1-\frac{t}{t_\text{decay}}\right)^{2/3}\right. &  -(2-\beta)\left(1-\frac{t}{t_\text{decay}}\right)^{4/3}\nonumber \\ 
		&\left.  -(\beta-1)\left(1-\frac{t}{t_\text{decay}}\right)^{2}\right]. \label{Eq__pme2t}	
\end{align}
\par The modification mostly affect our results in the limit $r\lesssim r_0$, or equivalently $t\ll t_\text{decay}$, as expected. Both $S^{(3)}(r)$ and $S^{(3)}(t)$ are shown in Fig.~\ref{f-Pcs--Modified} (green) alongside those obtained from $n=2$ replicas (orange) and those of Page (blue) for comparison. The consistency between the results from $n=3$ replicas and those of Page is obvious. Please note that for $\beta=1$ the modified results coincide with the old ones. It is worth noting that the modified curves (Fig.~\ref{f-Pcs--Modified}, green) might be considered as smooth versions of the Page curves up to some $\mathcal{O}(r^8)$, or equivalently $\mathcal{O}((t_\text{decay}-t)^{8/3})$, corrections.
\par We can draw some general conclusions regarding the contributing terms to the entanglement entropy. According to Eq.~\eqref{Eq__S3}, these terms take the form
\begin{equation}
	-\frac{\theta_n r^{2n}}{2nG^n},
\end{equation}
where $n=1,2,3,\ldots$. The coefficients, $\theta_n$'s, are determined by the underlying gravitational theory. As indicated in subsection \ref{Subsec__Interpret_In}, these coefficients are positive for $n>1$, while $\theta_1$ is negative. Although the connection between the series and $G$ is not immediately obvious---due to the presence of $G$ raised to various powers in the coefficients (see Eq.~\eqref{Eq__new-consts})---it is evident that this series does not emerge from a perturbative expansion in $G$. Indeed, we do not expect to derive a smooth Page curve from a perturbative treatment of $G$. Notably, the series consists exclusively of even powers of $r$. This characteristic arises from our \nth{1} holographic assumption combined with the replica trick, whereby each replica contributes a term proportional to the area, or $r^2$. Additionally, the signs of the coefficients are subject to constraints imposed by the consistency requirements detailed in Sec.~\ref{Sec__Shape_Dependence}. As indicated in that section, over-calculations in the first term must cancel out in the higher-order terms. Furthermore, the \nth{2} holographic assumption aids in determining the lowest coefficient, $\theta_1$. Hence, it can be contended that the assumptions we made and the consistency constraints we imposed have been pivotal in shaping the ultimate outcome, Eq.~\eqref{Eq__pme2r}.
\par In this section, we considered the entanglement entropy, $S$, as a function of the black hole radius, $r$. Although there might be much more quantities with complicated functionalities present in the problem, as we only consider the variation of the entanglement entropy with time, and there is a one-to-one correspondence between the time and the black hole radius (see Eq.~\eqref{Eq__MtM0} with $M=r/2G$), we considered $S$ as an explicit function of $r$, while taking other quantities implicitly into account through the constants that appear in the series (namely Eq.~\eqref{Eq__S3}). It should also be noted that our approach is not fundamental. We made use of some assumptions such as: The entanglement entropy will not exceed the Bekenstein-Hawking entropy, $A/4G$---At most, it saturates the bound. Such assumptions should be generic when treating the problem using a fundamental approach. Also the complicated quantities should explicitly enter when using a fundamental treatment.
%%% Numerical Analysis %%%%%%%
\subsection{Numerical support}\label{Sec__Num_Ana}
%%%%%%%%%
The condition $\mathcal{O}(10^{-1})$ of $A_0/4G$, we explained in paragraph prior to section \ref{Sec__higher_replicas}, is met for $r/r_0 \in \mathcal{R}=[0, 0.35) \cup (0.95, 1]$. It is known that the Page curve describes the evaporation process more accurately at this region, whereas its performance is less reliable at the middle region. A numerical work on the data extracted from $S(r)$ of the Page curve in $\mathcal{R}$ region, might test our green curve of Fig.~\ref{f-Pcs--Modified}. We may choose some points in $\mathcal{R}$, and fit a polynomial of degree 6 to these points. A normal fit uses square error (SE)
\begin{equation} \label{SE}
	\text{SE} = \sum_j \left(S_\text{fit}(r_j) - S_\text{Page}(r_j)\right)^2,
\end{equation}
to fix coefficients of the polynomial, where $j$ runs over all selected $r_j/r_0$'s in the range $\mathcal{R}$. But this curve crosses the Page curve several times that is not accepted by the principle of fine-grained entropy being always $\leq$ coarse-grained entropy.
\par
To keep the fit below the page curve, we should use a constrained fit. This constraint can be implemented by modifying the minimization process. Instead of solely minimizing the square error (SE), we minimize the following quantity
\begin{equation}
	\text{SE}+\lambda\sum_{i}\left(\max\{0,S_\text{fit}^\text{c}(r_i)-S_\text{Page}(r_i)\}\right)^2,
\end{equation}
where the superscript $\text{c}$ indicates the constrained fit, and $i$ runs over all $r_i/r_0$'s in the range $[0,1]$. In this formulation, the first term represents the ordinary square error of relation (\ref{SE}), and the second term introduces a penalty that applies whenever the fitted curve crosses the Page curve. By setting $\lambda$, the Lagrange multiplier, to a sufficiently large value (e.g., of $\mathcal{O}(10^6)$), the minimization process incurs a substantial penalty if the fit exceeds the Page curve. This effectively encourages the fit to stay below the Page curve, even if it means increasing the square error. In scenarios where the fit remains below the Page curve, the second term becomes zero, allowing the minimization to focus solely on reducing the square error. This approach guarantees that the resulting curve provides the best possible fit while adhering to the essential constraint of staying below or equal to the Page curve.
\begin{figure}[t]
	\centering
	\includegraphics[width=\linewidth]{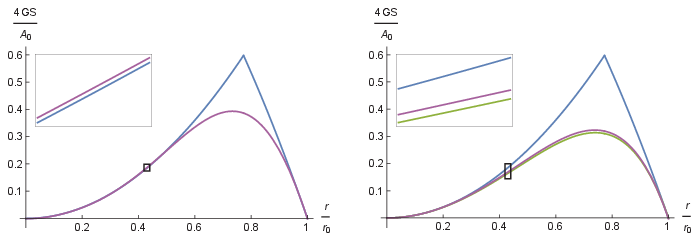}
	\caption{\label{f-fit-1} Left: A normal fitted polynomial of degree 6 (purple) to the Page curve (blue) are shown. This normal fit exceeds the Page curve in certain regions, indicating that it does not accurately represent the fine-grained entropy. Right: A constrained fit of degree 6 polynomial (purple) to the Page curve (blue), and the curve from $n=3$ replicas (green) are shown. This fit always stays below the Page curve.}
\end{figure}
\par
Figure \ref{f-fit-1} shows results of the two scenarios. In the left subfigure, the fitted polynomial is shown in purple alongside the Page curve in blue. At the box on top of this subfigure, we showed a crossing event more closely in a small range. We see that the polynomial may cross the Page curve, that is not allowed. In the right figure, the constrained fit in purple is shown alongside the $n=3$ replicas curve in green and the Page curve in blue. At the box on top of this subfigure we showed that the fitted polynomial stays below the Page curve at the same range. In fact it stays below the Page curve in all the domain, in agreement to the principle of fine-grained entropy being always $\leq$ coarse-grained entropy. The consistency between the fitted curve and the $n=3$ replicas curve is evident. The relative error is consistently less than 3.1\%, indicating a strong agreement between the curves.
\section{Conclusions}\label{Sec__Summ}
%%%%%%%%%%
The Page curve exhibits a sharp turnaround arising from the phase transition between a vanishing quantum extremal surface and a non-vanishing one. This study introduced a method for obtaining a smooth version of the Page curve which corresponds to a gradual transition. Comparing the original and smooth curves, we obtained that a gradual transition can lead to a substantial increase ($\sim 50\%$) in the informational content of the emitted radiation.
\par Naturally, the entanglement entropy ($S$) is expected to depend on the system's boundary ($\chi(x)$), making it a functional denoted as $S(\chi)$. In order to ensure the well-defined nature of the entanglement entropy (single-valued), it should remain unchanged when the boundary is initially deformed from $\chi$ to $\chi + \delta\chi$ and then returned to $\chi$. Based on this insight and employing expansions under the assumption of smoothness, certain constraints on the entanglement entropy were derived. These constraints played a crucial role in determining the desired curve for the entanglement entropy. It is important to mention that some holographic assumptions were used to match the limiting behaviors of the obtained curve with those of Page. Thus, the obtained curve can be considered as a smooth version of the Page curve.
\par It is worth noting that the mathematical identities derived in this work, coupled with the obtained interpretations, exhibit a remarkable generality. They remain applicable across various scenarios, provided the crucial smoothness condition is met.
\par We investigated the simplest case of a Schwarzschild black hole in flat spacetime. However, the consistency constraints and holographic assumptions we explored can also be applied to more complex black hole configurations, such as Kerr black holes or those present in asymptotically Anti-de Sitter (AdS) spacetime. In the context of a general black hole characterized by mass, angular momentum, and charge, the evaporation process results in the gradual loss of these properties, which must be considered in calculations. We anticipate that incorporating the latter two quantities will add complexity to the calculations and results. Additionally, various asymptotic behaviors of spacetime introduce effects that should be considered. For example, in the case of an evaporating black hole in AdS spacetime, the radiation may return from the boundaries to the black hole. Consequently, sufficiently large black holes could reach a stationary state where the rate of emission matches the influx of returning radiation from the boundaries, preventing complete evaporation. In this scenario, it is crucial to consider this influx in our calculations, as we anticipate that it will result in the entanglement entropy converging to a limiting value over extended timescales.
\section*{Acknowledgments}
We would like to express our appreciation to Don N.~Page for his helpful comments and to Tarek Anous for his valuable tips.
%%%%%%%%%%%%%%%%%%%%%%%%%%%%%%%%%%%%%%%%%%%%%%%%%%
\providecommand{\href}[2]{#2}
\begingroup\raggedright
\providecommand{\href}[2]{#2}\begingroup\raggedright
%\bibliographystyle{ws-ijmpd}
%\bibliography{ijmpd-2.bib}
\endgroup
\end{document}